\documentclass{article}

\usepackage{arxiv}

\usepackage[utf8]{inputenc} 
\usepackage[T1]{fontenc}    
\usepackage{hyperref}       
\usepackage{url}            
\usepackage{booktabs}       
\usepackage{amsfonts}       
\usepackage{nicefrac}       
\usepackage{microtype}      
\usepackage{lipsum}		
\usepackage{graphicx}
\usepackage[numbers]{natbib}
\usepackage{doi}
\usepackage{caption}
\usepackage{subcaption}
\usepackage{amsmath}

\title{Network based Approach Estimating COVID-19 Spread Patterns}

\author{ 
{
\hspace{1mm}Jiarui Dong}\\
	School of Science\\
	Beijing University of Posts and Telecommunications\\
	Beijing 100876, China\\
	\texttt{jiaruidongdse0808@gmail.com.com} \\
	\And
    {
    \hspace{1mm}Guanghao Ran} \\
	School of Science\\
	Beijing University of Posts and Telecommunications\\
	Beijing 100876, China \\
	\texttt{ranguanghaonuaa@163.com} \\
}

\renewcommand{\shorttitle}{\textit{arXiv} Template}

\hypersetup{
pdftitle={A template for the arxiv style},
pdfsubject={q-bio.NC, q-bio.QM},
pdfauthor={David S.~Hippocampus, Elias D.~Striatum},
pdfkeywords={Contagion, Synchronization Networks, Signal Propagation},
}

\begin{document}
\maketitle

\begin{abstract}
    In this study, we construct a series of evolving epidemic networks by measuring the correlations of daily COVID-19 cases time series among 3,105 counties in the United States. Remarkably, through quantitative analysis of the spatial distribution of these entities in different networks, we identify four typical patterns of COVID-19 transmission in the United States from March 2020 to February 2023. The onsets and wanes of these patterns are closely associated with significant events in the COVID-19 timeline. 
    Furthermore, we conduct in-depth qualitative and quantitative research on the spread of the epidemic at the county and state levels, tracing and analyzing the evolution and characteristics of specific propagation pathways.
    Overall, our research breaks away from traditional infectious disease models and provides a macroscopic perspective on the evolution in epidemic transmission patterns. This highlights the remarkable potential of utilizing complex network methods for macroscopic studies of infectious diseases.
\end{abstract}


\section{Introduction}
Despite the spread of COVID-19 seems to reach a halt in 2023, this rampant pandemic has caused disastrous aftermaths in the past three years. Till the day of 10th March 2023, when Johns Hopkins University stopped collecting COVID-19 data, the total cases globally surpassed 678 million with death toll exceeding 6.8 million. Thus, it would never be late for people to retrospect the process of suppressing the transmission of the pandemic. In this article, we attempted to explore the fundamental principles behind the spread of COVID-19 in a novel perspective.

We gathered the daily data on increase case of 3105 counties around the US, the length of data ranging from January 2020, when the first case spotted in 
Washington to March 2023.  Manipulating these time series data, we then established a series of synchronization networks, the techniques that have already been put into use in climatology\cite{mondal2023global}.

Based on these networks, we separate them into several patterns according to the distributions of \textit{Divergence} and EMD distances among the networks. To figure out how virus is spread in different patterns and inspired by methods in \cite{chennubhotla2007signal}, we explore the virus propagation in a scale of \textbf{state} and also assume it a discrete-time Markov process. By this means, we can calculate the expected hitting time between two states and find its relation to network topologies.  

In addition, we also refer our theoretical analysis to the measures taken by governments and events arisen during the period we study so as to figure out how they influence the transmission of the virus and how the virus urge people to take corresponding actions in reverse.

The structure of the paper would be arranged as follows. The experimental results and the detailed analysis are described in Section \ref{sec:Results}. Section \ref{sec:Discussion} concludes the methodologies we perform in this study and clarify our preponderances over existing models, shortcomings and the follow-up research plan. Finally, in Section \ref{sec:Methods}, the comprehensive explanation of the research method will be listed.

\section{Results}
\label{sec:Results}
\begin{figure}[htbp]
    \centering
    \includegraphics[scale=0.5]{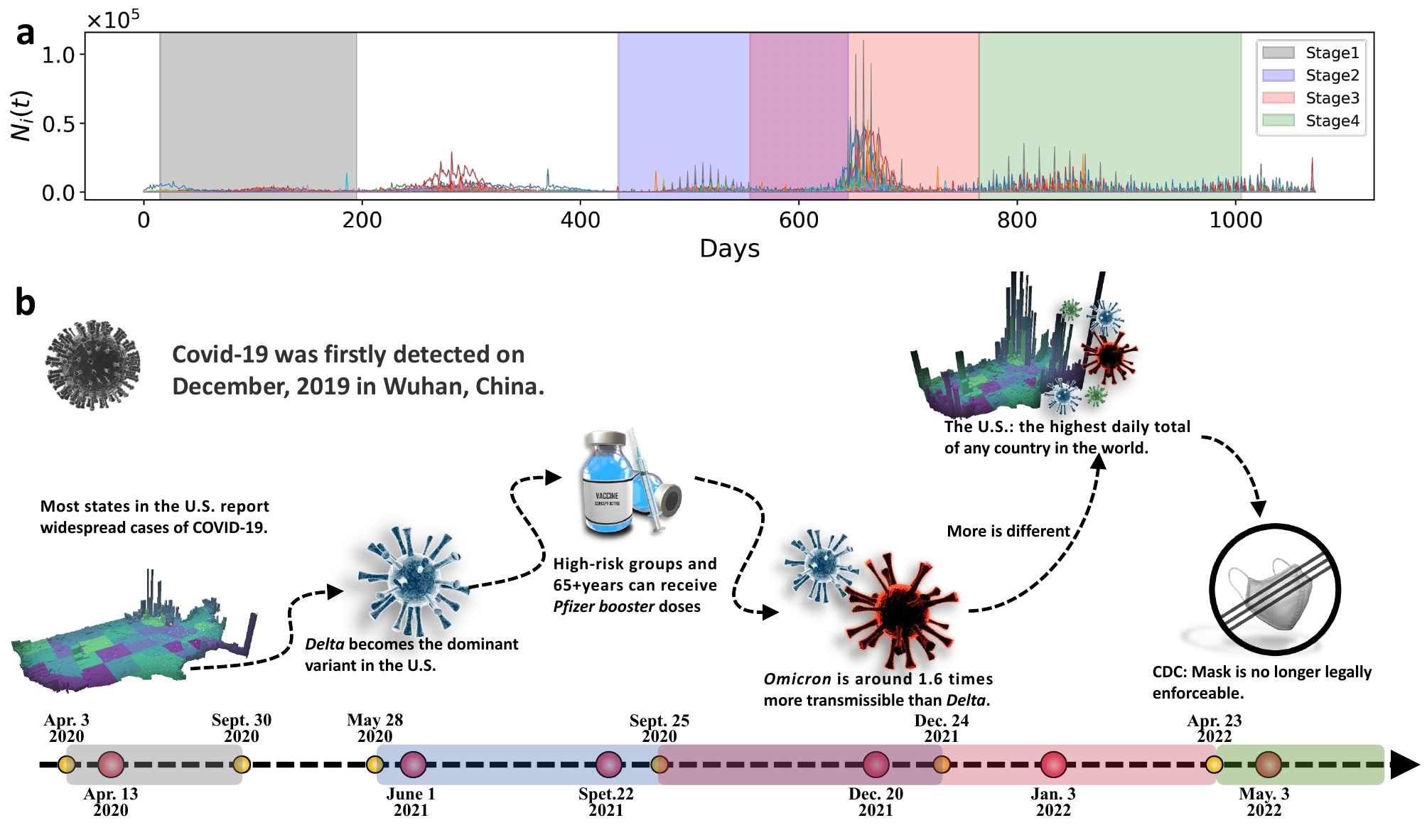}
    \caption{\textbf{(a) The time series of daily new cases in 3105 counties.} The period spans from March 19, 2020 to Feb. 27, 2023. The duration of each stage is shaded by corresponding color marked in \textbf{(a)}.  \textbf{(b) Timeline of the Covid-19 pandemic in the U.S. since its widespread across the country with illustrations of the significant moments reported by CDC}. The red dots on the timeline represent the dates of significant events reported by the CDC. The yellow dots denote the start and end dates of each stage in the study. Different stages are distinguished by the colors aligned with what are marked in \textbf{(a)}.}
    \label{fig:timeline}
\end{figure}

We collect daily cumulative COVID-19 case data from 3,105 counties in the United States from March 19, 2020, to February 17, 2023, totaling 1,065 days, as reported in The New York Times. Based on this data, we compute the daily new case time series for each county $i$, , denoted as $N_{i}(t)$, as shown in Fig. \ref{fig:timeline}{\bf a}.

To investigate the asynchronous spread of the virus between counties, we employ the Pearson time-delay cross-correlation function (CCF), as mathematically described in Section \ref{sec:methods}, Eqs.\ref{ccf1} and \ref{ccf2}. This method has been widely used in climate and environmental research, revealing global climate teleconnection propagation pathways \cite{liu2023teleconnections, meng2023arctic} and tracing the pathways of pollutant dispersion\cite{vlachogiannis2021correlation}. Thus, we seek to investigate the patterns of COVID-19 transmission from a novel perspective of complex networks.

However, considering the rapid outbreak of the virus and the diversity of human behavioral patterns, it proves challenging to identify stable transmission patterns within short time windows. Consequently, in this study, we determine a time series length of approximately 120 days (roughly 4 months) as the window for calculating the CCFs. Moreover, given that the estimated incubation period of COVID-19 is up to approximately 15 days\cite{lauer2020incubation, linton2020incubation}, we set the time delay $\tau$ in Eqs.\ref{ccf1} and \ref{ccf2} to range from -15 to 15 days.

Based on these settings, we treat the 3,105 counties as nodes in the network and compute the CCF between each pair of nodes (i.e., Node $i$ and Node $j$). The highest peak of the CCF, $\rho(\theta_{ij})$, is assigned as the weight of the corresponding link, with the sign of $\theta_{ij}$ indicating the direction of the link. Furthermore, to study the spatiotemporal evolution characteristics of the virus network, we perform a sliding window analysis with a 30-day (1-month) interval, updating the starting point of the time series. This procedure result in a total of 32 networks.

It is worth noting that through this process, we obtain a series of fully connected networks. However, some links with relatively small or even negative weights are not statistically significant. Therefore, in each network, we retain only the maximum 5\% of links based on their weights.

\subsection{Network Perspective of COVID-19 Millstones} \label{sec:patterns}

In the realm of complex network theory, the degree centrality of network nodes stands as a fundamental and pivotal characteristic, employed to measure the level of connectivity of a node in the network\cite{boccaletti2006complex, albert2002statistical}. 
In this research endeavor, we delineate a novel concept termed "Divergence" (Eq.\ref{div}) for nodes, which indicates the difference between out-degree and in-degree. The primary objective is to quantitatively assess the role, e.g., {\it Broadcaster} or {\it Receiver}, played by a node in the propagation of an epidemic.  Consequently, within the aforementioned ensemble of 32 networks, each characterized by distinct topological structures, we compute the Divergence for each individual Node $i$ out of the total of 3,105 nodes, denoted as $Div_i(k)$, where $k=1,2,...,32$. Subsequently, we discern an intriguing pattern: for certain neighboring networks, the spatial distribution of Divergence exhibits striking similarity. This observation suggests that the internal structure of these networks remains relatively stable throughout the designated time frame, implying a degree of consistency in the patterns of epidemic spreading.

\begin{figure}[htbp]
    \centering
    \includegraphics[scale=0.5]{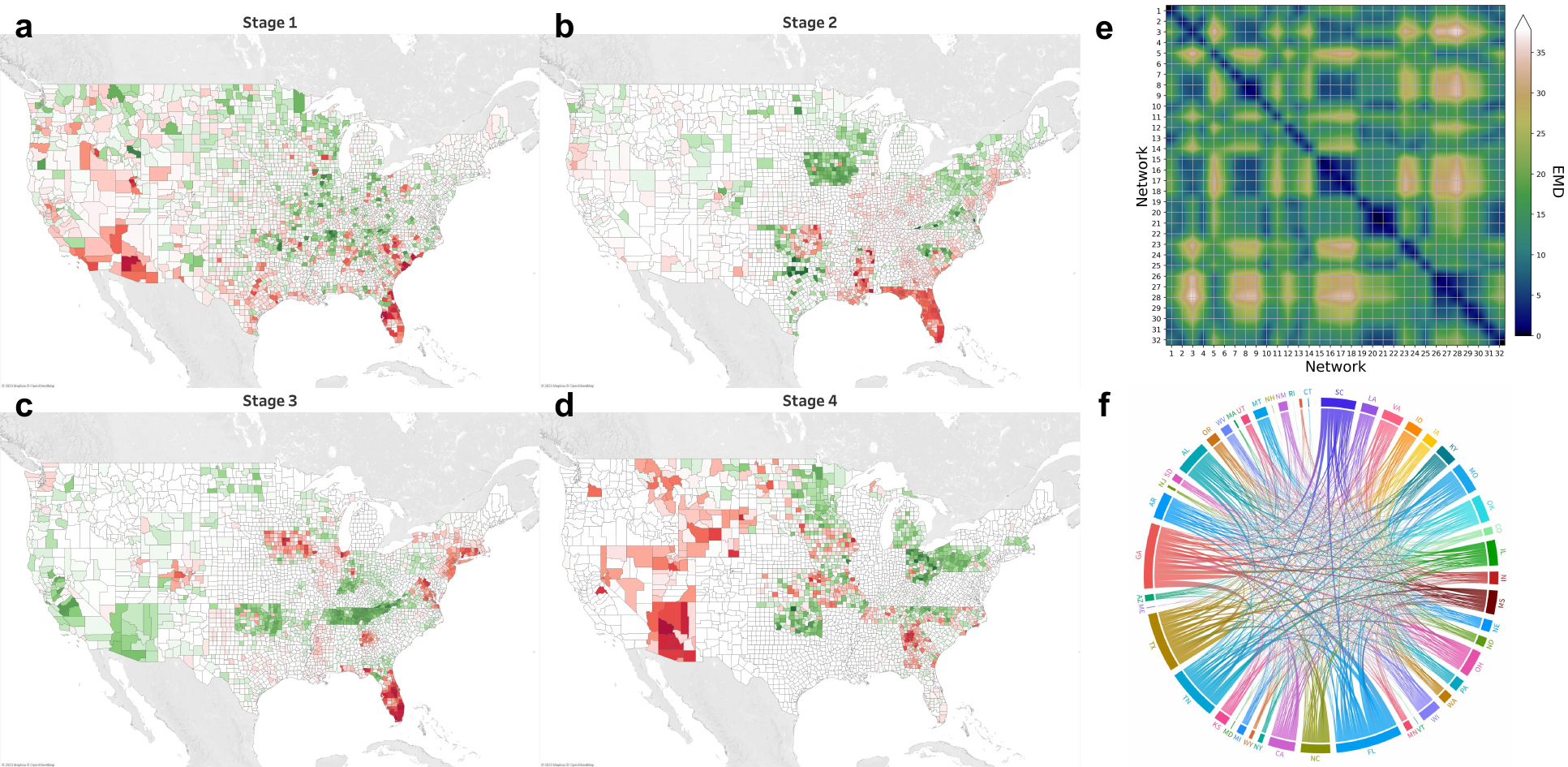}
    \caption{\textbf{(a)-(d) County divergence distributions of 4 stages.} Counties colored as red are broadcasters, the green ones are receivers and the left white ones are neutral. \textbf{(d) The Earth Mover's Distances among different networks.} The networks separated into the same stage differs way less than those are in the different stage from them. \textbf{(f) The chord map manifesting the influence flow among all 48 contiguous States} The arc represents both the influence volume inserted on State B by State A and that inserted on State A by State B (see the different thickness of one arc's two ends.)}
    \label{fig:pattern}
\end{figure}

To quantitatively explicate this phenomenon, we introduce the EMD distance, commonly known as the Earth Mover's Distance (EMD), which quantifies the minimal cost required to transform one distribution into another\cite{ramdas2017EMD}, and be widely applied in comparing the similarity between two images. Thus, we calculate the EMD between each pair of the 32 spatial distributions (networks), as illustrated in Fig.\ref{fig:pattern}{\bf e}. Given the focus on the progressive evolution of networks, our analysis exclusively considers temporally adjacent networks. Thus, our attention is concentrated on the main diagonal of the EMD matrix. 

Consequently, this examination reveals distinct stages, whereby the 32 networks can be discontinuously divided into 6 distinctive stages. Of these stages, we particularly concentrate on 4 stages, namely, Apr. 3, 2020 to Sept. 30, 2021, May 28, 2021 to Dec. 24, 2021, Sept. 25, 2021 to Apr. 23, 2022, and Apr. 23, 2022 to Dec. 19, 2022. In Fig.\ref{fig:timeline}{\bf a} and {\bf b}, these time periods are represented by gray, blue, red, and green intervals respectively. Furthermore, Fig.\ref{fig:pattern}{\bf a-d} illustrates the spatial distribution maps of the average network divergence for each of the 4 stages.

Strikingly, we have observed a strong alignment between these time periods and significant milestones in the COVID-19 pandemic, as reported by the Centers for Disease Control and Prevention (CDC). Specifically, on April 10, 2020, the United States surpassed Italy and Spain to become the global epicenter of the pandemic, experiencing the highest number of COVID-19 cases and fatalities. Merely three days later, on April 13, 2020, the majority of states in the U.S. reported widespread community transmission of COVID-19. Fig.\ref{fig:pattern}{\bf a} illustrates the spatial distribution of the network node divergence during this time period. Qualitatively, it is evident that the initial outbreak of COVID-19 in the United States spread from coastal cities to inland regions. Among these, California, Texas, and Florida exhibited higher transmission capabilities compared to other states, which can be closely associated with their larger population sizes (the three most populous states in the U.S. which can be referred to from the website \url{https://worldpopulationreview.com/states}) and geographic positioning (coastal cities with higher population mobility\cite{lee2020human, kang2020multiscale}).

Furthermore, on June 1, 2021, the Delta variant (B.1.617.2), initially identified in India, emerged as the dominant variant in the United States, signifying an escalating impact of variant strains on the U.S. epidemic. This event marked the beginning of the second stage, as depicted in Fig\ref{fig:pattern}{\bf b}. Subsequently, on September 22, the U.S. Food and Drug Administration (FDA) authorized booster doses of the Pfizer vaccine, specifically targeting individuals aged 65 and above, as well as younger adults at high risk of severe COVID-19 or with frequent exposure to the virus. This event corresponds to the onset of the third stage, as shown in Fig\ref{fig:pattern}{\bf c}.

In December 2021, the U.S. reported the first cases of the Omicron variant (B.1.1.529), which was estimated to have approximately 1.6 times higher transmissibility compared to the Delta variant, according to CDC data released on December 20. This indicates that as the Omicron variant gained dominance, the second stage was coming to an end. One reason for this transition is that the Delta variant still maintains significant protection against currently available COVID-19 vaccines\cite{bian2021impact, shiehzadegan2021analysis}, while the Omicron variant, with its numerous mutations, has the potential to attenuate vaccine neutralization capacity and diminish protective efficacy\cite{hoffmann2022omicron, wilhelm2021reduced}. From Figs.\ref{fig:pattern}{\bf b} and {\bf c}, we can observe that Florida consistently exhibited a high transmission capability during both stages. Additionally, states along the northeastern coast, such as Maryland, Pennsylvania, New York, and Massachusetts, emerged as significant broadcasters during this period. However, the distribution of heavily impacted states varied, likely due to the different transmission speeds of the Delta and Omicron variants\cite{allen2023comparative, ren2022omicron}.

Looking ahead, on January 3, 2023, the United States witnessed an astonishing milestone as it reported nearly one million new COVID-19 infections in a single day, setting a global record for the highest daily surge in cases. Alarmingly, within just one week, hospitalizations due to COVID-19 surged by nearly 50\%. On May 3, 2023, the Centers for Disease Control and Prevention (CDC) issued a recommendation for the continued use of masks indoors in transportation hubs as a preventive measure against COVID-19 transmission, although this guidance no longer carried legal mandates.

Fig.\ref{fig:timeline}{\bf a} illustrates that during the third stage, the daily number of new cases experienced an incredible outbreak followed by a gradual return to stability. However, the implementation of this policy relaxation led to a minor peak in daily new cases, indicating a transition from the third stage to the fourth stage.

Notably, these time points are identified through the measurement of network similarity, and they have been empirically validated to exhibit a strong and intrinsic correlation with major real-world events. Consequently, we contend that the correlation network constructed based on daily new cases data effectively captures the dynamic variations in the transmission patterns of the epidemic, attributable to factors encompassing viral mutational dynamics, vaccine interventions, and public health management policies.

\subsection{Epidemic Spreading Paths}
In Fig.\ref{fig:pattern} {\bf a-c}, we find that Florida, located in the southeastern corner of the United States, plays the role of a broadcaster in the first three stages of the epidemic. This is largely attributed to the fact that Florida has the third-largest population in the United States, following California and Texas. Due to its geographical location and significant population growth and mobility, it is not surprising that Florida acts as a "pioneer" in the spread of the epidemic. 

However, we are curious about how Florida, as a "pioneer", transmits the epidemic to other regions. Therefore, we need to identify the "paths" in the networks within different stages. Although the spread of the epidemic does not strictly follow a single route, for simplicity, we focus on finding the most probable path. Considering that we have constructed a directed weighted sparse network, the Dijkstra algorithm\cite{dijkstra2022note} becomes our preferred choice.

Firstly, we define the expense of each existing path, denoted as $E_{ij}$, as the reciprocal of the weight of the directed link, i.e.,
\begin{equation}
    E_{ij} = \frac{1}{\rho(\theta_{ij})}.
    \label{Expense}
\end{equation}

Additionally, considering the local scope of population movement, we believe that even if a path has a small expense, it is not logical if the corresponding real-world distance is significantly large. Therefore, we impose a threshold $\mathcal{D}$ on the real geographic distance for the path search algorithm.

\begin{figure}[htbp]
    \centering
    \includegraphics[scale=0.6]{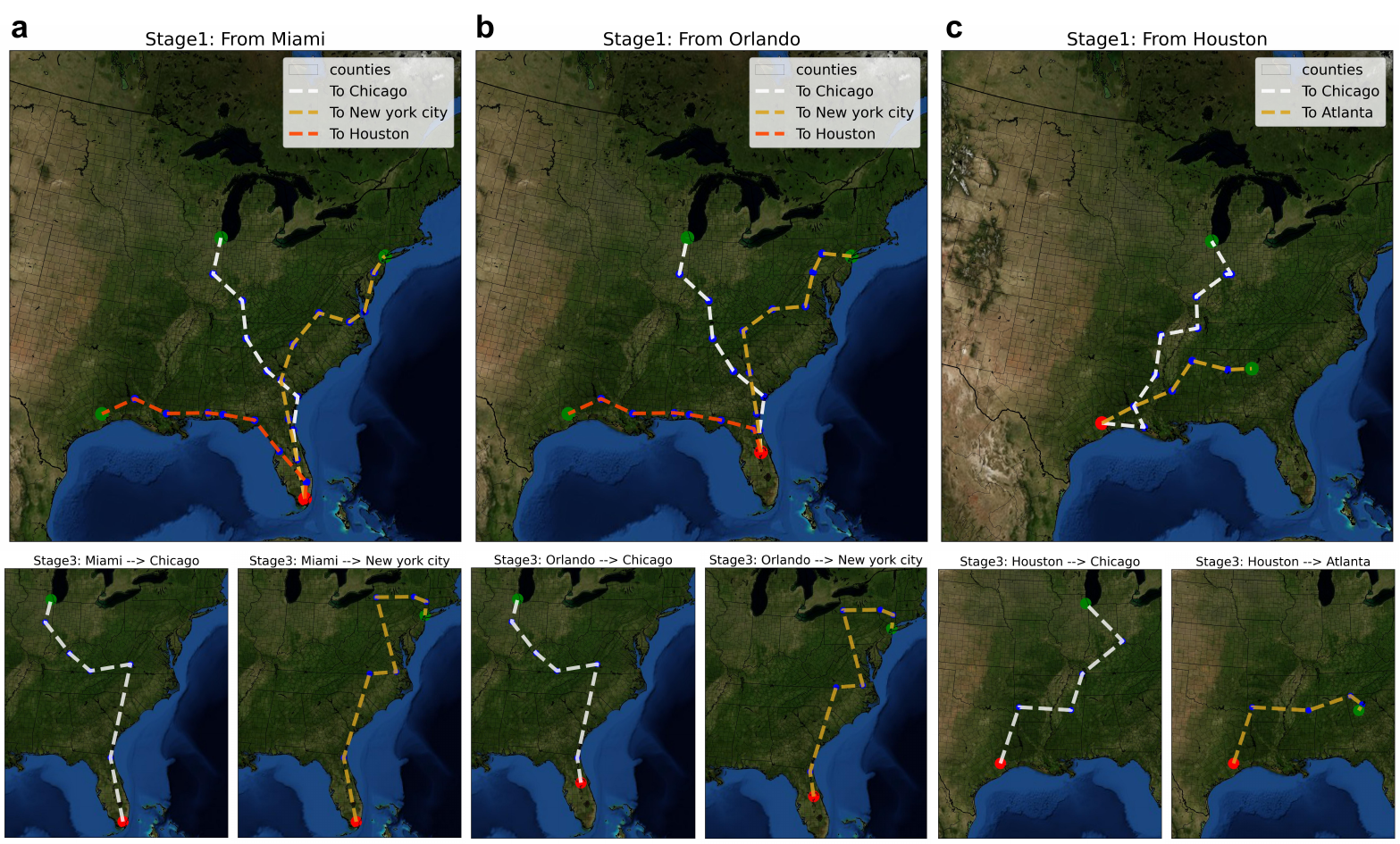}
    \caption{\textbf{The examples of the shortest paths with restrictions on geographical distance between certain significant counties in different Stages.} The geographical distance between two counties is set less than $300 Km$ in Stage 1 while in Stage 3 it is set less than $800 Km$ in \textbf{(a) , (b)}, and $600 Km$ in \textbf{(c)}. The looser restriction allows more intermediate nodes in network to select. \textbf{(a)} The shortest paths from Miami to Chicago, New York City and Houston in Stage 1 and the shortest paths from Miami to Chicago and New York city in Stage 3. \textbf{(b)} The shortest paths from Orlando to Chicago, New York City and Houston in Stage 1 and the shortest paths from Miami to Chicago and New York city in Stage 3. \textbf{(c)} The shortest paths from Houston to Chicago and Atlanta in Stage 1 and the shortest paths from Houston to Chicago and Atlanta in Stage 3. The shortest paths from Miami to Chicago and New York City and from Houston to Chicago and Atlanta pass less intermediate counties in Stage 3 than in Stage 1.}
    \label{fig:path}
\end{figure}

In the initial stage, our exploration begin from Miami, the capital city of Florida, and encompassed three distinct routes leading to Houston, Chicago, and New York City. A distance threshold of $\mathcal{D}=300 Km$ is chosen based on the predominant modes of human transportation within this range, which include walking, cycling, automobiles, and railways\cite{barbosa2018human}. As illustrated in Fig.\ref{fig:path}{\bf a}, the paths to these destinations are delineated using orange, white, and yellow dashed lines. The orange path follows the westward route of Interstate 10, traversing Alabama, Mississippi, and Louisiana before reaching Houston in Texas. The white path extends northward, spanning Georgia, Tennessee, and Kentucky to ultimately arrive at Chicago in Illinois. The yellow path overlaps with the white path in Georgia and proceeds northeastward, passing through North Carolina, Virginia, Maryland, and New Jersey, eventually reaching New York City in New York State. Moreover, we also discover for paths from Orlando to the aforementioned destinations, as Orlando is located in the center of Florida and has more highways connected to it compared to Miami. The results are shown in Fig.\ref{fig:path}{\bf b}.

\label{paths}

In order to scrutinize the different propagation patterns between different stages, our analysis focuses on the fixed starting points of Miami and Orlando and the endpoints of Chicago and New York City during the Stage 3. However, it is observed that no viable propagation paths from Miami or Orlando to these cities existed when applying a distance threshold of $\mathcal{D}=300 Km$. This finding suggests that not only did the network structure undergo transformations during this stage, but there were also significant changes in the actual geographical distances between each connection. Only when the distance threshold is increased to $\mathcal{D}=800 Km$, propagation paths from Miami to Chicago and New York City are identified, as depicted in Fig.\ref{fig:path}. Within this distance range, specifically $300 Km<D\leq 800 Km$, the primary modes of human transportation shifted to railways, airplanes, and ships\cite{barbosa2018human}.

These observations indicate that our network reveals the changing propagation patterns in central cities of Florida between these stages. It is found that in the early stages of the outbreak, the propagation paths aligned with the interstate highway routes in the United States. However, in the most intense phase of the outbreak, the connections between Florida and neighboring states disappeared, giving way to longer-distance propagation paths.

In addition to Florida, we also consider another populous state, Texas. In Texas, our focus is primarily on exploring the economic hub of the state , Houston, and the propagation paths from Houston to the transportation hub cities of Chicago and Atlanta in the western United States. Similarly, in Fig.\ref{fig:path}{\bf c}, we observe similar phenomena as before, and these patterns are also found in the propagation paths starting from Houston.

\subsection{Virus Propagation}
In Section \ref{sec:patterns}, we have already showed the patterns we identified and the characteristics of them. Additionally, in this part, we will delve into the propagation of virus from a source to a target. However, in the last part, we only examine our results at the scale of county, while in this part we will explore the virus propagation at the scale of {\bf states}. Specifically, when it comes to {\bf states}, the {\bf state} with many counties being broadcasters in the last section may change its role, since some counties just influence its near neighbors within the same {\bf state}, without extending the impact on other {\bf states}.

To figure out where one {\bf state's} total impact flows, we first reform our networks. In the previously study, each node in the networks represent a county. Now, we try to carry out a simple renormalization to incorporate all of the counties within a {\bf state} into a new node, representing a {\bf state} in the new networks. Then we calculate the impact for each pair of {\bf states}, denoted as {\bf state} $i$ and {\bf state} $j$. For each county in {\bf state} $i$, we aggregate its impact on all the counties in {\bf state} $j$ and sum the aggregations up to get the influence of {\bf state} $i$ on $j$. As a county will insert impact on counties of the same {\bf state} as itself, in our new networks, the entries on the diagonal will not be all zeroes. This provides us a intuitive way to find out the {\bf states} barely impact {\bf states} other than itself but the impact only circulates inside. 

Based on the new networks, we can investigate how virus propagates from a source {\bf state} to a target {\bf state}. First, we notate the adjacency matrices of the networks as $\mathbf{A}=\{a_{ij}\}$, and for {\bf state} $i$ in a network, its degree $d_i$ can be
\begin{equation}
     d_i = \sum_{j=1}^{n}a_{ij},
\end{equation}specifically $\mathbf{D}=diag\{d_{i}\}$. Therefore, we assume the process as a Markov process and extract stochastic matrices from the new networks. Details are shown in Sec. \ref{sec:stochastic}.

As our main interest is on the mechanisms of the virus' diffusion from {\bf state} to {\bf state}, in other words, the signal propagation across the network, we refer to the work of Chennubhotla et el\cite{chennubhotla2007signal}, which looked into the communications between proteins under the assumption of Gaussian network models. Next, we introduced the concept of expected hitting time, which is important when it comes to Markov process and the commute time appeared in \cite{chennubhotla2007signal}. The expected hitting time $H(i,j)$ could be defined as the expected time till we hit {\bf state} $j$ starting from {\bf state} $i$, while according to \cite{chennubhotla2007signal}, commute time $C(i,j)$ will be
\begin{equation}
    C(i,j) = H(i,j) + H(j,i) = C(j,i).
\end{equation}
Borrow the results from the methods in \cite{chennubhotla2007signal} and the calculation in Sec. \ref{sec:hitting}, the component form of Eq.\ref{(8)} can be realized as
\begin{equation}
    H(i,j)=\sum_{k=1}^{n}\left\{\mathbf{\Gamma}^{-1}_{ki}- \mathbf{\Gamma}^{-1}_{ji}-\mathbf{\Gamma}^{-1}_{kj}+\mathbf{\Gamma}^{-1}_{jj}\right\}d_k.
\end{equation}

Due to the existence of loops in the new network, it would be tricky to calculate the inverse of the Laplacian matrix, as the determinant of it is quite small. In this literature, we use the pseudo-inverse of the Laplacian matrix to approximate its genuine inverse. Besides, the numerical precision and scaling also contribute to the inaccuracies of calculation. Thus these factors may lead that summation of matrices in Fig. 5 not necessarily accurately equals to the matrix in Fig. 4(a). However, the physical meanings of each term decomposed from the Laplacian matrix remain valid.

Thus, the laplacian matrix can be decomposed into three factors, the target term $\mathbf{\Gamma}_{jj}^{-1}$, the intermediate term $\mathbf{\Gamma}_{ki}^{-1}-\mathbf{\Gamma}_{kj}^{-1}$ and the source-target term $-\mathbf{\Gamma}^{-1}_{ji}$. 
We illustrate the expected hitting time matrix and commute time matrix of \textbf{stage} 1 in Fig. 4(a) and Fig. 4(b). We notice that it is more variant in terms of the columns of matrix $H$ than the rows. Namely, the {\bf states} are more variant in capacity when they are as {\it receivers} than they are as {\it broadcasters}. These phenomena can also be observed in Fig. 4(c) and Fig. 4(d) and we notice the {\bf states} that not only possess relatively small $ \left<H_{bro}(i,j) \right>$ but also fairly large $\left<H_{rec}(i,j) \right>$. Specifically, these {\bf states} are effective in transmitting virus while unproductive in receiving. 

Additionally, to figure out the factors influencing the expected hitting times, we show the distributions of target term, source-target term and intermediate term in Fig. 5(a), Fig. 5(b) and Fig. 5(c). We discover that target term only make negative contributions to elongating the expected hitting times while the contributions made by target-source term and intermediate term are indeterminate. Besides, compared with the target terms, the scale of intermediate terms and target-source terms is negligible. These distributions can exactly explain the phenomena we see in Fig. 4(c)-(d) and Fig. 3. In Fig. 3, we find as for the same kind of transmission path (i.e. geodesics) between 2 \textbf{states} in \textbf{stage} 1 and \textbf{stage} 3, the amount of the nodes interpolating in the paths become less as the \textbf{stage} processes. Referring to the results shown in Fig. 5(a)-(c), for the negligible contribution provided by the intermediate terms compared with the target terms, we can provide an potential explanation to this phenomenon, as the source and target are static, the presence of the intermediate nodes in the paths makes little difference to the transmission. Additionally, in Fig .4(c)-(d), the obvious discrepancy in the capacity of receiving and the uniformity in the capacity of broadcasting verify our conclusion that the modulating effect of the sources are also minor in comparison with the targets. That's to say, the target terms play a dominant role in determining the expected hitting times, and the sources can adjust the effect of the target terms, whereas the effect of intermediates and sources are insignificant.

If we add up $H(i,j)$ and $H(j, i)$, we can have the commute time
\begin{equation}
    C(i,j)=C(j,i)=(\mathbf{\Gamma}^{-1}_{ii}+\mathbf{\Gamma}^{-1}_{jj}-\mathbf{\Gamma}^{-1}_{ij}-\mathbf{\Gamma}^{-1}_{ji})\sum_{k=1}^{n}d_k,
\end{equation}
the source-target terms $\mathbf{\Gamma}^{-1}_{ij}$ and $\mathbf{\Gamma}^{-1}_{ji}$  are not combined together due to the asymmetry of $\mathbf{\Gamma} ^{-1}$. In the sense of resistance distance\cite{klein2002resistance}, we have
\begin{equation}
    C(i,j)=C(j,i)=\mathbf{\Omega}_{ij}\sum_{k=1}^{n}d_k.
\end{equation}
Resistance distance measures the difficulty for signal flowing from one node to another in a network. In this way, we can assume the commute time as the cost the virus needs to pay when transmitting between two \textbf{states}. Assuming the velocity the virus transmits from one \textbf{state} to another along the edge in our network is always static equals to $\mathbf{1}$, we show the correlations of the commute distance with both geographical distance and flight volume in Fig. 5(d)-(e). In Fig. 5(e), we find there is no significant correlations between commute distance and geographical distance, as geographical distances are uniformly distributed irrelevant of the commute distances. This manifests the proximity does not effectively determine the commute distance when the virus spread in the network. In contrast, the flight volume show negative correlations with commute distance. When the air traffics between two \textbf{states} are busy, the commute distance tends to become shorter. As $C(i,j)$ is the summation of $H(i,j)$ and $H(j,i)$, if we suppose \textbf{state} \textit{i} and \textbf{state} \textit{j} are targets of each other, we provide an evidence for the evaluation of the effects of target terms. This sheds light that the transmission of virus can overcomes the barrier set by proximity (e.g. via long-range transportation) and it would be more efficient to contain the spread by considering the exact circumstances of different target \textbf{states}.

\begin{figure}
\hspace*{-1cm}
    \centering
    \includegraphics[width=\textwidth]{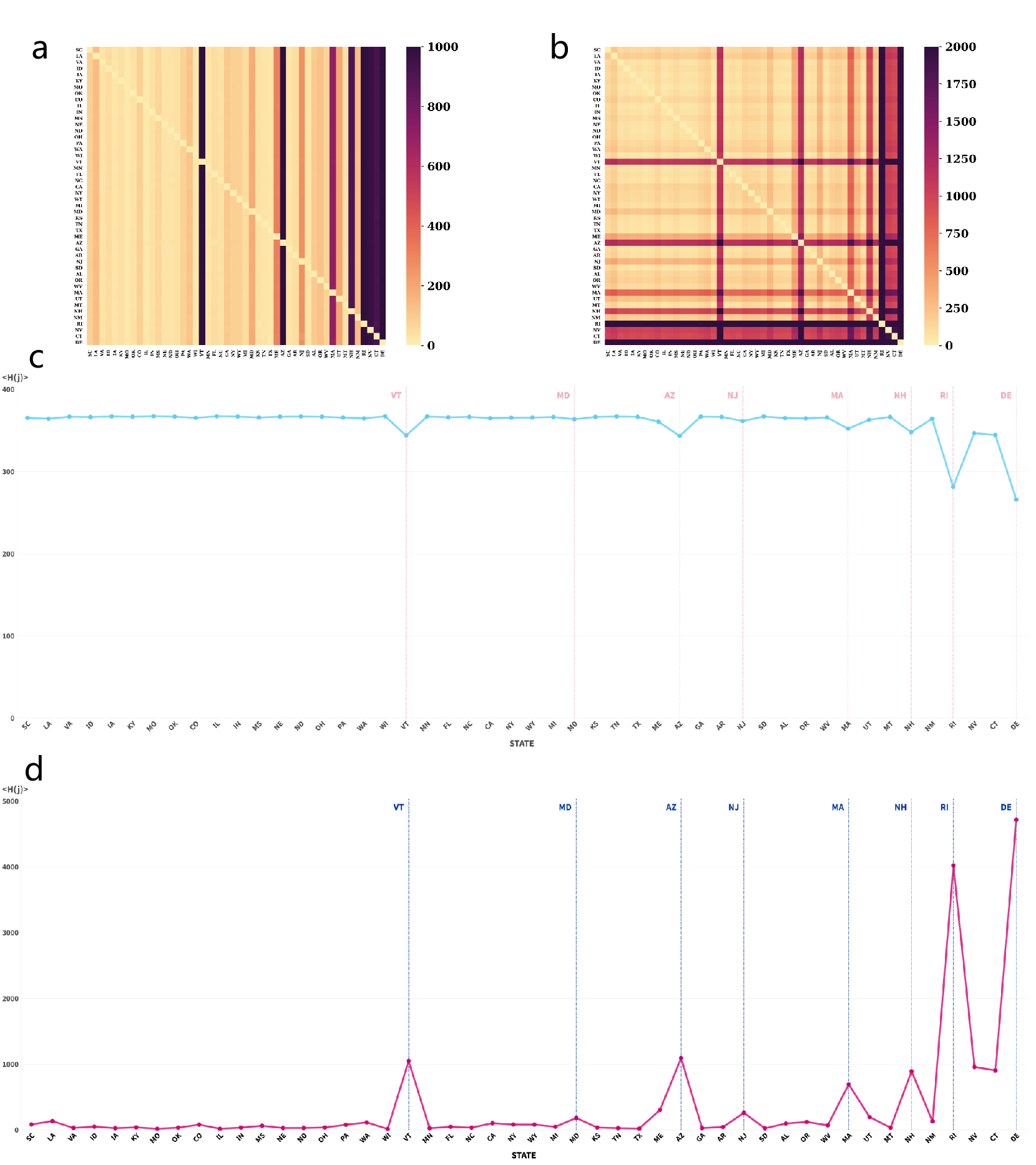}
    \caption{\textbf{(a)-(b) Distribution of expected hitting times and commute times in Stage 2.} \textbf{(a)} $H(i,j)$ denotes the expected hitting time between {\bf state} \textit{A} to {\bf state} \textit{B}. Rows denote the processes of broadcasting and columns denote the processes of receiving. \textbf{(b)} $C(i,j)$ represents the expected commute time between {\bf state} \textit{A} and {\bf state} \textit{B}. \textbf{(c)-(d) The average expected hitting time of each {\bf state} when it is as a broadcaster and as a receiver.} The average of each row $ \left<H(i) \right>$ is of less standard variance (78.5) compares to that (1091.9) of the average of each column $ \left<H(j) \right>$ of $H$, manifesting {\bf states} in Stage 2 are of similar capacity in propagating influence while possess different inclinations in receiving impacts. The {\bf states} of low $ \left<H(i) \right>$ and high $ \left<H(j) \right>$ are marked with dash lines in \textbf{(c)} and \textbf{(d)} respectively.}
    \label{fig:hit}
\end{figure}

\begin{figure}[htpb]
\hspace*{-1.25cm}
    \centering
    \includegraphics[width=\textwidth]{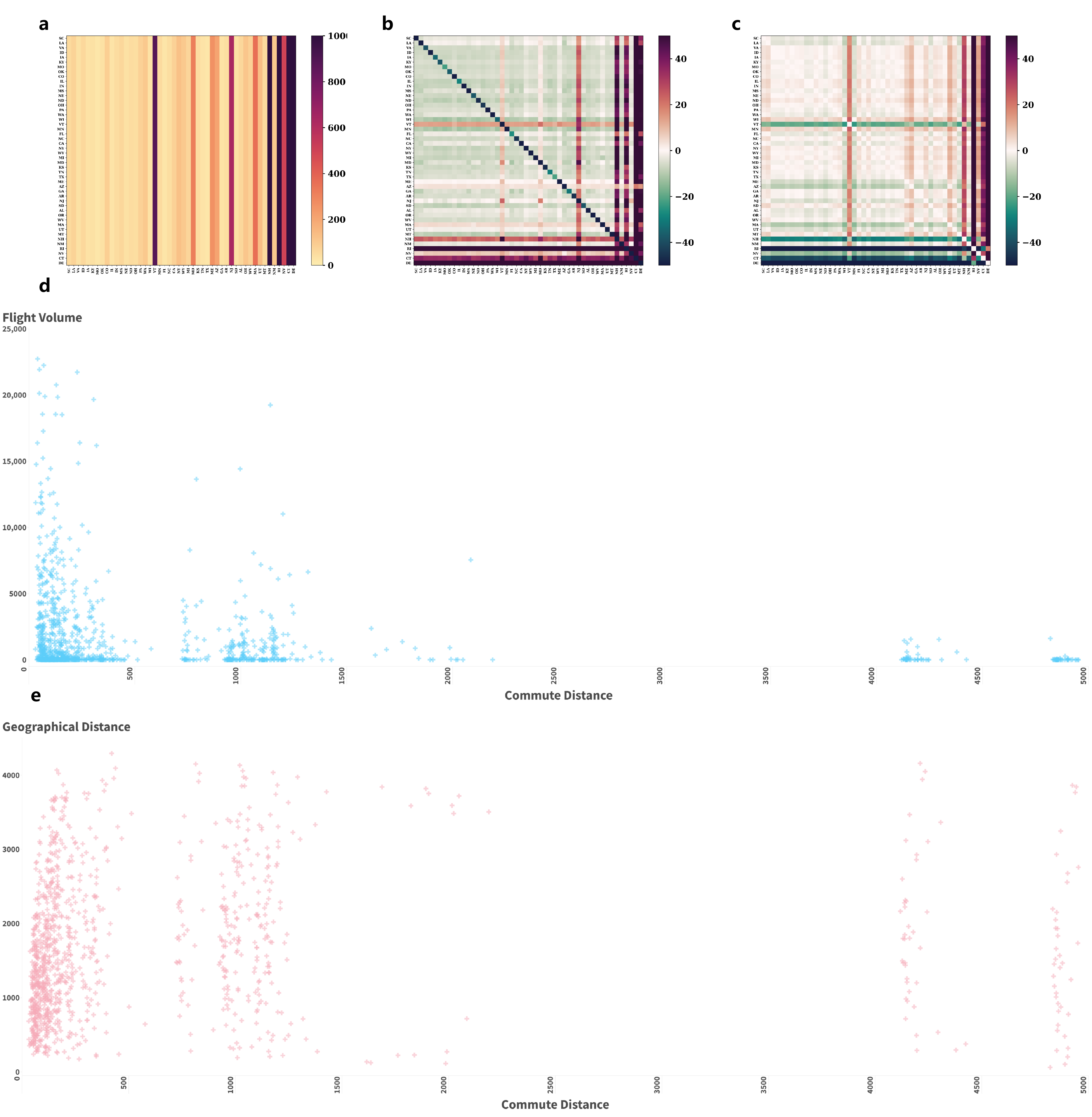}
    \caption{\textbf{Decomposition of the Laplacian matrix of the network of {\bf states} in Stage 2, where (a) target term $\mathbf{\Gamma}_{jj}^{-1}$, (b) source-target term $-\mathbf{\Gamma}_{ji}^{-1}$ and (c) intermediate term $\mathbf{\Gamma}_{ki}^{-1}-\mathbf{\Gamma}_{kj}^{-1}$.} 
    Due to the self loops of the state network in Stage 2, the Laplacian matrix is near singular. The $\mathbf{\Gamma^{-1}}$ here is approached with the pseudo-inverse of the Laplacian matrix. Although the decompositons cannot necessarily completely reproduce the matrix in Fig. 4(a), the physical meaning of each terms remains valid. 
    \textbf{(a)} The target terms are only relevant with the properties of the target {\bf state} and only make contributions to elongating the expected hitting time, which is in accordance with what is discovered in Fig. 4(c)-(d). As for the one entity term, the target plays an more important role than the source. \textbf{(b)} The source-target terms manifests the intervention of the sources in determining the expected hitting time among {\bf states}. Nonetheless, their contributions are indeterminate and minor comparing with the target terms. \textbf{(c)} The intermediate terms also make indeterminate contributions to expected hitting time. However, compares with the contributions of the target terms shown in \textbf{(a)}, the intermediate term contributions are also negligible, which aligns with what we discovered in Fig. 3. \textbf{(d)-(e)} Correlations of commute distance with interstate flight volume in the period of \textbf{stage 1} and with geographical distance.}
    \label{fig:laplace}
\end{figure}

\clearpage
\section{Discussion}
\label{sec:Discussion}
To better comprehend the mechanisms underlying the transmission of COVID-19 virus, we put forward a set of novel methodologies to investigate the spread patterns of it across the US contiguous states and the formations of them. Specifically, we performed a network based approach to reach this aim. We collect the data of the daily increase case of 3105 counties in the US contiguous states since the date when the first case was detected in the US mainland till Feb. 26, 2023. With these data, we calculate the time-delayed cross-correlation functions (i.e. \textit{CCF}) any two counties. By assuming the counties as the nodes and the \textit{CCF}s as the edge weights in the networks and sliding the time window for the calculation of \textit{CCF}s, we can establish a series of consecutive directed networks along time, which represent the interactions among states and the propagation of virus. We also propose a metric named as \textit{Divergence} to measure the influence influx and outflow of a specific county. According to the observation of the distributions of the \textit{Divergence}s in every network and the \textit{EMD} distance among networks, we separate the networks into four groups and call them \textit{stage}, namely, the patterns. We find the separation of patterns are not only due to the mathematical and structural resemblance of the networks within but there are also real-world incidents recorded by the US Centers for Disease Control and Prevention supporting our judgements. For clarity, we underscored the shortest paths between several metropolitans in different \textbf{stages}. In addition to analysis in the scale of county, we also extend our study to the scale of \textbf{state}, since more phenomena and mechanisms might be unveiled from a different perspective. We renormalize the former county node in networks into \textbf{state} node and recalculate the edge weights between two \textbf{states}. With these new networks and supposing virus transmission a Markov process, we can bridge the network structure and the expected hitting times the virus need to spend from one \textbf{state} to another. Furthermore, we can decompose the Laplacian matrices of the networks and extract the effect of each factor (i.e. targets, sources and intermediates) in virus propagation. And we spot targets matter the most when virus transmitting, while sources and intermediates can only slightly modulate the effect of targets. We also notice that geographical distances between \textbf{states} cannot influence virus transmission across the network, whereas heavy air traffics can significantly reduce the cost of transmission. These findings can exact correspond to what we discovered in the study carried out in county scale and offer reasonable explanations to it. 

Overall, our study have some edges over the existing research. For one thing, we do not model individuals that carrying virus and also impose no specific assumptions on them like SIS model. This can prevent us from inappropriate evaluation and preconception about the characteristics of the virus and the behaviors of the hosts. In contrast, we establish our models solely out of phenomena we observe, to wit, data-driven. Besides, we introduce the real-world incidents and various metrics (in this literature, \textit{Divergence} and \textit{EMD} distance) to verify our model to make it more convincing and reasonable. For another, we implement our study in various scales, say county and \textbf{state}, allowing us to look into the mechanisms from a different perspective. What's more, some phenomena can only be witnessed in certain scale. For instance, some county may be efficient \textit{broadcaster} in county scale, however, it may just influence county closely next to itself and become less significant when we observe from the \textbf{state} scale. Last but not least, we propose a hybrid approach to elucidate the formation of the spread patterns. By presuming the virus transmission a Markov process, we can extend the network structure to the Markov chain and thus endow the physical meanings with the factors we decomposed from the Laplacian matrices.

However, it is worth noting that there are still quite a few limitations to our research and flaws also exist. First off, Pearson correlation coefficients are one metric measuring the correlation with two time series and there may be further correlations between them undetected by Pearson correlation coefficients. Besides that, the switch of scale may lose some information and due to the network structure and numerical errors, the accuracy of calculation is not completely guaranteed.  Lastly, our models are tend to be descriptive rather than predictive, therefore, the criticality for the transition between patterns is not proposed.

In conclusion, at the time when the COVID-19 virus seems to cease to spread, we take the US contiguous states for an example and retrospect the progress of the epidemic in the past 2 years and bring up new models for the comprehension of COVID-19 transmission, and our study illuminates for the further knowledge of COVID-19 pandemic. This pandemic has caused fatal disaster to people's health and world economy due to our ill preparation for health emergency, consequently, it is of great importance to conclude experience from the past and get adequately ready for the upcoming challenge.


\section{Date and Methods}
\label{sec:Methods}

\subsection{Data Collection}
\subsubsection{U.S. Daily cases data}
The data repository used in this study is derived from The New York Times' GitHub repository (\url{https://github.com/nytimes/covid-19-data}). The primary data published in this repository are the daily cumulative number of cases and deaths reported in each county and state across the U.S. since the beginning of the pandemic.

In this study, we calculate the daily new cases based on the daily cumulative number of cases ($T_{i}(t)$), 
\begin{equation}
    N_{i}(t) = T_{i}(t) - T_{i}(t-1).
\end{equation}
Considering our focus on studying the spread of the pandemic, we exclude some counties that are geographically distant from the mainland U.S., including Alaska, Hawaii, Puerto Rico, Northern Mariana Islands, and Virgin Islands. Therefore, out of the original 3,223 counties, we retain 3,105 counties located within the mainland U.S., denoted as $i=1, 2, ..., 3105$.

On the other hand, we find that many counties had no recorded cases during the early stages of the outbreak, which pose challenges in measuring the similarity in time series. Therefore, we set March 19, 2020, as the starting point for the time series of new cases in all counties, and the endpoint is set as February 17, 2023. This is because as of March 2023, The New York Times announced that the data for daily cases and deaths would no longer be updated. Hence, for all 3,105 counties in the U.S., a total of $1,065 \times 3,105$ sets of observations are aggregated in the prepared dataset for further exploration.

\subsubsection{Interstate flight volume}
We collect data from the FAA (Federation Aviation Administration) Operations \& Performance Data website (\url{https://aspm.faa.gov/}). The data incorporates the flight count between airports in a designated period of time. For instance, when calculating the fight count between \textbf{state} \textit{i} and  \textbf{state} \textit{j}, we first find out all the airports that belong to \textbf{state} \textit{i} and \textit{j} respectively, and then sum up the flight counts between all the airport pairs.

\subsection{Networks Construction}

\subsubsection{Links}
To define links between each pair of node $i$ and node $j$, we follow\cite{yamasaki2008climate} and compute the time-delayed, cross-correlation function
\begin{equation}
    \rho^k_{ij}(\tau) = \frac{\langle N^k_{i}(t)N^k_{j}(t+\tau)\rangle - \langle N^k_{i}(t)\rangle \langle N^k_{j}(t+\tau)\rangle}
    { \sigma\left[N^k_{i}(t)\right] \sigma\left[N^k_{j}(t+\tau)\right]}, \tau<0,
    \label{ccf1}
\end{equation}
and
\begin{equation}
    \rho^k_{ij}(\tau) = \frac{\langle N^k_{i}(t-\tau)N^k_{j}(t)\rangle - \langle N^k_{i}(t-\tau)\rangle \langle N^k_{j}(t)\rangle}
    { \sigma\left[N^k_{i}(t-\tau)\right] \sigma\left[N^k_{j}(t)\right]}, \tau\geq0
    \label{ccf2}
\end{equation}
Here, $k$ represents the $k$th network, and the angle brackets denote the average over consecutive days. The variable $\tau \in [-\tau_{\max}, \tau_{\max}]$ represents the time lags.

Specifically, we identify the highest peak, $\max\limits_{\tau}(\rho^k_{ij}(\tau))$, in the cross-correlation function, refered as to the weights of the link between node $i$ and node $j$ in the networks. Meanwhile, we denote the corresponding time lags as $\theta^k_{ij}$, i.e., $\max(\rho^k_{ij}(\tau))=\rho(\theta^k_{ij})$. Moreover, the signs of the time lags indicate the direction of each link. Specifically, a link from node $i$ to node $j$ is established when the time lag is positive ($\theta^k_{ij} \geq 0$). 

Consequently, the network is represented by an anti-symmetric adjacency matrix denoted as $\mathbf{A(k)}$. In this matrix, the entry $a_{ij}(k)$ denotes the presence of a link from node $i$ to node $j$, assigned a value of $1$, while a link from node $j$ to node $i$ is represented by $-1$.

\subsubsection{Node Degree}
 Our particular focus lies on the degree centrality of nodes, a fundamental parameter in network theory that quantifies the total number of connections for each node \cite{albert2002statistical}. Each node is characterized by two distinct degrees: in-degree and out-degree. The in-degree represents the average strength of incoming links, whereas the out-degree indicates the average strength of outgoing links. Mathematically, these degrees can be defined as follows:
\begin{equation}
    IND_{i}(k) = \sum_{j=1,j\neq i}^{M} \rho(\theta^k_{ij})I_{a_{ij}(k)=-1},
    \label{in-degree}
\end{equation}
and
\begin{equation}
    OUTD_{j}(k) = \sum_{i=1,i\neq j}^{M} \rho(\theta^k_{ij})I_{a_{ij}(k)=1},
    \label{out-degree}
\end{equation}
where $M$ equals to the total number of nodes (i.e., 3,105) and $I$ is the indicator function. Based on the equations above, we define the "Divergence" of each node $i$ as:
\begin{equation}
    Div_i(k) = OUTD_i(k) - IND_i(k).
    \label{div}
\end{equation}
\label{sec:methods}

\subsection{Signal propagation in networks}
\subsubsection{Derivation of stochastic matrix}
\label{sec:stochastic}
We derive the stochastic matrices out of the new networks where nodes are assumed as \textbf{states}, namely, the transmission of virus is treated as a Markov process. 
We notice that for some {\bf states} the entries located on the diagonal of the stochastic matrix are close to 1. In this sense, these {\bf states} could be deemed as nodes disconnected with the large connected component in the network or absorbing {\bf states} in a Markov process. 
After specifying the adjacent matrix $\textbf{A}=\left\{a_{ij} \right\}$ and the degree matrix $\textbf{D}=diag\left\{d_{i} \right\}$, we define the probability that {\bf state} $i$ gives the impact to {\bf state} $j$ in single step
\begin{equation}
    p_{ij} =  \frac{a_{ij}}{d_{i}}.
\end{equation}
These {\bf states} will cause the stochastic matrix to be reducible, thus make it trivial for us to calculate expected hitting time of a certain {\bf state}. Therefore, for simplicity, we set up a threshold to eliminate them to ensure the stochastic matrices are irreducible and the expected hitting time is not infinity. Notably, after elimination, the stochastic matrices have to be normalized again, making sure the summation of every row equals 1 (i.e., $ \sum_{j=1}^{n}p_{ij}=1$) . Note that our networks are all directed, in this study, our Laplacian matrix is \textit{out-degree Laplacian matrix}.  
We can easily deduce the relation between the stochastic matrix and adjacency matrix
\begin{equation}\label{(4)}
    \mathbf{P}=\mathbf{AD^{-1}}.
\end{equation} 
\subsubsection{Expected hitting times}
\label{sec:hitting}
Assume we have a stochastic matrix $\mathbf{P}$ $\in \mathbf{R}^{n\times n}$, where $n$ is the total number of {\bf state}, and we intend to calculate $\eta_{ij}$, the expected hitting time from \textbf{state} $i$ to \textbf{state} $j$, we can have equations
\begin{equation}
    \mathbf{\hat{I}}-\mathbf{\hat{P}}_j=\mathbf{\hat{1}},
\end{equation}
where $\mathbf{\hat{I}}$ is an identity matrix of $n-1$ dimension, $\mathbf{\hat{P}}_j$ is the very stochastic matrix but with $j^{th}$ row and column eliminated and $\mathbf{\hat{1}}$ is a column vector of length $n-1$ and all element 1. Solve the equation above we get a vector representing $\eta_{ij}$, i can be $1,2,...,j-1,...,n$, namely $\mathbf{\hat{H}}_j$, we have
\begin{equation}
     \mathbf{\hat{H}}_j=\mathbf{\hat{1}}+ \mathbf{\hat{H}}_j\mathbf{\hat{P}}_j.
\end{equation}
With $\mathbf{\Gamma } = \mathbf{D}-\mathbf{A}$ and Eq. \ref{(4)} and with Laplacian matrix truncating $j^{th}$ row and column, we derive
\begin{equation}\label{(8)}
    \mathbf{\hat{H}}_j=\mathbf{\hat{1}}\mathbf{\hat{D}}\mathbf{\hat{\Gamma }}^{-1},
\end{equation}
where $\mathbf{\hat{D}}$ and $\mathbf{\hat{\Gamma}}$ are truncated as $\mathbf{\hat{P}}_j$ likewise.

\bibliographystyle{unsrt}
\bibliography{references}  

\end{document}